\documentclass[conference]{IEEEtran}
\IEEEoverridecommandlockouts
\usepackage{cite}
\usepackage{amsmath,amssymb,amsfonts}
\usepackage{graphicx}
\usepackage{textcomp}
\usepackage{xcolor}
\def\BibTeX{{\rm B\kern-.05em{\sc i\kern-.025em b}\kern-.08em
    T\kern-.1667em\lower.7ex\hbox{E}\kern-.125emX}}

\usepackage{amsmath,amssymb}
\usepackage{booktabs}
\usepackage{cite}
\usepackage{graphicx}
\usepackage{multirow}
\usepackage{array}
\usepackage{tabularx}
\usepackage{makecell}
\usepackage{xcolor}
\usepackage{xspace}
\usepackage{url}
\usepackage{stfloats}
\usepackage{balance}
\usepackage{booktabs,tabularx}
\usepackage{tikz}
\usetikzlibrary{arrows.meta,positioning,calc}
\usepackage{graphicx}
\usepackage{listings}

\newcommand{\system}{{\textsc{TabClean}}\xspace}

\newcommand{\dirty}{\mathcal{D}}
\newcommand{\clean}{\mathcal{C}}
\newcommand{\devset}{\mathcal{A}}
\newcommand{\prog}{\mathcal{P}}
\newcommand{\guard}{\mathsf{g}}
\newcommand{\Guard}{\mathcal{G}}
\newcommand{\schema}{\mathcal{S}}

\usepackage[ruled,vlined,linesnumbered]{algorithm2e}

\makeatletter
\providecommand{\linebreakand}{%
  \end{@IEEEauthorhalign}%
  \hfill\mbox{}\par%
  \mbox{}\hfill%
  \begin{@IEEEauthorhalign}%
}
\makeatother

\definecolor{tcDirty}{RGB}{174,70,49}
\definecolor{tcGuard}{RGB}{30,91,150}
\definecolor{tcTransform}{RGB}{27,125,77}
\lstdefinestyle{tabclean}{
  language=Python,
  basicstyle=\ttfamily\scriptsize,
  breaklines=true,
  columns=fullflexible,
  keepspaces=true,
  showstringspaces=false,
  frame=single,
  rulecolor=\color{black!20},
  xleftmargin=0.5em,
  xrightmargin=0.5em,
  aboveskip=0.35em,
  belowskip=0.15em,
  escapeinside={(*@}{@*)}
}

\begin{document}

\title{\system: Reusable LLM-Synthesized Programs for Tabular Data Cleaning}

\author{\IEEEauthorblockN{Yibo Wang}
\IEEEauthorblockA{\textit{Purdue University} \\
West Lafayette, USA \\
wang7342@purdue.edu}
\and
\IEEEauthorblockN{Riteng Zhang}
\IEEEauthorblockA{\textit{Purdue University} \\
West Lafayette, USA \\
zhang5982@purdue.edu}
\and
\IEEEauthorblockN{Yinghao He}
\IEEEauthorblockA{\textit{Purdue University} \\
West Lafayette, USA \\
he923@purdue.edu}
\linebreakand
\IEEEauthorblockN{Yongye Su}
\IEEEauthorblockA{\textit{Purdue University} \\
West Lafayette, USA \\
su311@purdue.edu}
\and
\IEEEauthorblockN{Bharat Bhargava}
\IEEEauthorblockA{\textit{Purdue University} \\
West Lafayette, USA \\
bbshail@purdue.edu}
\and
\IEEEauthorblockN{Chunwei Liu}
\IEEEauthorblockA{\textit{Purdue University} \\
West Lafayette, USA \\
chunwei@purdue.edu}
}

\maketitle

\begin{abstract}
Reliable analytics and machine-learning pipelines depend on clean tabular data, yet production tables often contain missing values, typographical errors, inconsistent formats, violated dependencies, unit mismatches, and ambiguous categorical values. Existing cleaning systems make different trade-offs. Constraint-based systems need experts to specify rules. Learning-based systems need labels or retraining. Recent LLM-based cleaners reduce setup effort, but many call an LLM on rows, cells, or repeated workflow steps, so their cost grows with table size and with every recurring batch.

We present \system, a model-training-free system that compiles LLM reasoning into reusable guarded cleaning programs. Given a dirty table and a small annotated development set, \system profiles table evidence, diagnoses repair mechanisms, synthesizes executable Python transformations, validates candidates with cell-level feedback, and commits the best program for reuse on schema-compatible batches. The key abstraction is an evidence-backed guarded repair clause. A deterministic transformation may fire only when its dirty pattern, target-negative condition, evidence support, and scope constraints are satisfied. Across six benchmarks, \system achieves high precision, improves F1 over representative rule-based, learning-based, and LLM-based baselines on five datasets, and substantially reduces recurring runtime and API cost by replacing repeated LLM inference with deterministic program execution.
\end{abstract}


\section{Introduction}
Tabular datasets underpin decision-making in science, business, and the public sector, and they remain a dominant substrate for training and evaluating machine learning models \cite{tabulardata, treemodel}. In practice, however, tables are frequently contaminated by missing values, typos, inconsistent formats, unit mismatches, out-of-range measurements, and schema-dependent violations caused by sensor glitches and human workflows \cite{datacleaning, datacleaningbook}. It is widely estimated that data scientists spend the majority of their time cleaning and pre-processing raw data before being able to analyze it \cite{detectingerrors, tamrblog, sigmodblog, liu2025variable}. For example, scientists at Massachusetts General Hospital (MGH), one of the largest hospitals in the US, spend $80\%$ of their time building and refining data pipelines that involve extensive data preparation and model tuning \cite{datacivilizer2}. Left unaddressed, such errors can propagate into downstream analytics, bias model training, and erode trust in automated pipelines \cite{ambiguity,activeclean}.

The database community has studied this problem for decades. Rule- and constraint-based systems use denial constraints, conditional functional dependencies, matching dependencies, or domain rules to detect and repair inconsistent records \cite{tax,holistic,scared,holoclean}. Learning-based systems reduce some of the rule-authoring burden by using user labels, weak supervision, transfer learning, or active selection \cite{activeclean,raha,holodetect,baran}. Despite their successes, these approaches often require substantial domain expertise and iterative engineering. Rules must be written and maintained per dataset, constraints may be unknown or brittle under distribution shift, and specialized models often fail to transfer across schemas without retraining.

Large language models (LLMs) expand the design space for tabular data cleaning because they can interpret heterogeneous table contexts, natural-language hints, and domain-specific cues without task-specific training. Recent systems have applied LLMs to data wrangling, preprocessing, retrieval-assisted repair, standardization, workflow generation, and end-to-end cleaning~\cite{llmwrangle,llm_processor,retclean,cleanagent,autodcworkflow,cocoon,gidcl}. Existing LLM-based cleaners largely follow two paradigms~\cite{survey}. The first is prompt-based cleaning, where an LLM is repeatedly invoked to detect, verify, or repair dirty values~\cite{autodcworkflow,cleanagent,llmwrangle,iterclean,llm_processor,retclean}. The second is task-adaptive fine-tuning, where smaller language models are trained on dataset-specific error distributions~\cite{gidcl,beaver}. These approaches lower the barrier to expressing cleaning intent, but they shift the main bottleneck from rule authoring to scalability and reliability. Prompt-based systems incur inference cost at row, cell, or chunk granularity and therefore become expensive on large tables or recurring batches. Fine-tuning-based systems reduce per-call cost but introduce a cold-start requirement for high-quality training data and often generalize poorly to new schemas. Both paradigms can also produce unsupported repairs when the model must reason over long, noisy, or weakly grounded contexts.

\noindent\textbf{Motivating example.}
Consider a recurring flight-status table collected from heterogeneous web sources \cite{flight}. In a small annotated sample, a time attribute may contain values such as \texttt{730}, \texttt{7:30}, and \texttt{07:30 AM}, while the clean table expects a canonical \texttt{07:30} representation. A prompt-based cleaner must repeatedly ask an LLM to inspect many cells or chunks whenever a new batch arrives. A learning-based system may not have enough data for training and may not know which should be the target format. Yet for a human programmer, the desired solution is simple. Write a guarded parser that recognizes supported time formats, normalizes them, and leaves already-canonical or ambiguous values unchanged.

This example highlights our key observation. Many tabular errors are not isolated value-editing problems, but recurring schema-level repair mechanisms. A natural
alternative is therefore to use LLMs to synthesize executable cleaning logic rather than to repair
every value directly.
Some recent systems move in this direction through code or workflow generation~\cite{llmclean,cocoon}. However, their scope remains limited. LLMClean focuses on Ontological Functional Dependencies (OFDs), while Cocoon mainly generates regular expressions and SQL \texttt{CASE WHEN} clauses, which are often too narrow to capture heterogeneous error mechanisms and rarely transfer across future tables sharing the same schema. 
Yet narrow scope is not the main challenge. Even with broader coverage, a single prompt may produce plausible cleaning code, but plausible code is insufficient for reliable data repair. The system must distinguish strong table evidence from weak correlations, avoid over-correction, recover from executable failures, preserve high-precision transformations, and stop refinement when additional edits begin to hurt precision. Moreover, the generated program must be explainable, extensible, and transferable. It should express general repair logic rather than hard-coded replacements from the development set.

To address these challenges, we present \system\footnote{Code and data are available at \url{https://github.com/Wangyibo321/TabClean}.}, a scalable and model-training-free data cleaning system that formulates LLM-assisted cleaning as evidence-grounded synthesis of reusable guarded programs. \system uses a finite-state multi-agent pipeline to profile a table, diagnose repair mechanisms, plan guarded repair clauses, synthesize Python code, execute the code on a small human-annotated development set, and refine it using cell-level feedback. Rather than hard-coding dataset-specific replacements, agents draw on a shared, schema-agnostic skills library and apply repairs only when they are grounded in observable data evidence. The final program is cached and reused for future tables with the same schema, amortizing LLM cost across batches and reducing large-scale cleaning to ordinary program execution. Across six standard benchmarks, \system improves cleaning quality over representative rule-based, learning-based, and LLM-based baselines on five of them, and remains practical on large tables such as the 200K-row \textit{tax} benchmark while substantially lowering runtime and API cost.

We make the following contributions.
\begin{itemize}
    \item We formulate tabular data cleaning as evidence-grounded synthesis of schema-reusable guarded repair programs, enabling cleaning logic to be cached, audited, and reapplied across future tables with the same schema (Section~\ref{sec:problem}).

    \item We design an execution-guided synthesis loop that separates LLM-proposed repair mechanisms from deterministic execution, validation, safety checking, and stopping (Sections~\ref{sec:system-overview} and \ref{sec:fsm}).

    \item We instantiate this abstraction with repair-skill contracts, evidence levels, adaptive development-set construction, role-aware memory, guarded transformations, and best-program tracking (Sections~\ref{sec:program-semantics}--\ref{sec:agents-memory}).

    \item We evaluate \system on six standard data-cleaning benchmarks against four representative baselines, measuring cleaning quality, runtime, LLM cost, annotation sensitivity, ablations, and model sensitivity (Section~\ref{sec:experiments}).
\end{itemize}

\section{Problem Setting}
\label{sec:problem}

\subsection{Data Model and Objective}

Let $\dirty$ be a dirty table with schema $\schema=\langle A_1,\ldots,A_m\rangle$ and $n$ rows. Let $\clean$ be the corresponding clean table with the same schema and row alignment. A cell $(i,j)$ is dirty when $\dirty_{ij}\neq\clean_{ij}$ under a canonical comparison function that normalizes harmless representation differences such as surrounding whitespace when appropriate. A repair system returns $\widehat{\clean}$ and is evaluated by cell-level true positives, false positives, and false negatives. Precision measures the fraction of changed cells that are correct repairs, recall measures the fraction of dirty cells repaired, and F1 is their harmonic mean.

\system receives the dirty table and a small annotated development set $\devset$. The development set can be produced by human annotation, by a sampled historical dirty-clean pair, or by any workflow that supplies trusted clean values for a small subset of rows. \system uses $\devset$ only for synthesis and validation. Final quality is measured on held-out rows. This setting mirrors operational cleaning. The data owner can inspect a sample, but cannot label a full production table.

The goal is to synthesize a program $\prog$ that preserves schema and row order while repairing supported errors.
\begin{equation}
\max_{\prog}\; F_1(\prog(\dirty_{test}),\clean_{test}).
\end{equation}
LLM calls occur during synthesis and refinement. Applying a validated program to the full table, or to later batches with the same schema, should require no LLM inference.

\subsection{Reusable Cleaning Programs}

Many table errors are mechanistic. A time column may contain two parseable formats. A state column may violate a city-state dependency in rare rows. A numeric attribute may mix units. A bibliographic source may use several aliases for the same venue. Once the mechanism is known, a program can apply it across all rows with guards that prevent changes outside the supported pattern.

This observation motivates the central abstraction in \system. A repair program is a sequence of guarded repair clauses
\begin{equation}
\prog = \langle (s_1, \guard_1),\ldots,(s_k,\guard_k)\rangle,
\end{equation}
where $s_i$ is a repair skill, such as date normalization or dependency-based majority repair, and $\guard_i$ is a predicate that determines when the skill may fire. A clause is valid only when the table profile and development-set feedback provide evidence for both the repair and the guard. The same dirty value may therefore be repaired in one schema and left unchanged in another schema if the supporting evidence differs.

\subsection{Error Taxonomy}

\system structures diagnosis with an explicit taxonomy. The current catalog includes typographical errors (T), missing values (MV), format inconsistencies (FI), violated attribute dependencies (VAD), out-of-distribution values (OD), semantic equivalence or canonicalization issues (SE), unit and scale inconsistencies (UT), truncation or extraction artifacts (TE), schema and type inconsistencies (SI), and composite or misaligned fields (CM). The taxonomy is not meant to be complete. Its role is to force each proposed repair to state the evidence it requires. For example, an MV repair needs a deterministic source for the missing value, while an SE repair needs evidence that several spellings denote the same entity.

\section{System Overview}
\label{sec:system-overview}

\begin{figure*}[t]
  \centering
  \includegraphics[scale=0.34]{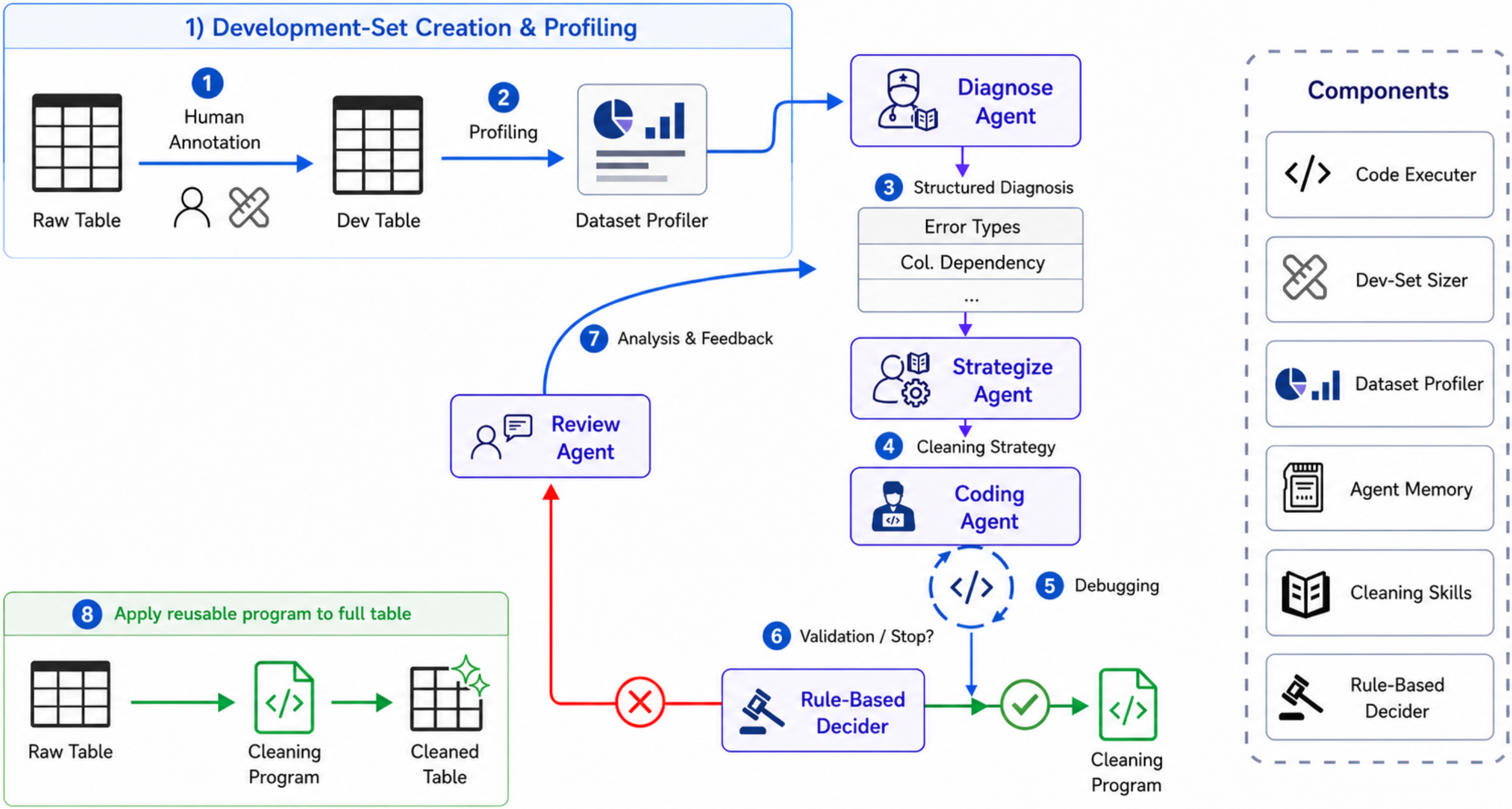}
  \caption{\system \ System Overview}
  \label{fig:workflow}  
\end{figure*}

Figure~\ref{fig:workflow} shows the end-to-end architecture of \system. \system follows a generate-validate-deploy design. LLM agents synthesize reusable cleaning logic from a small annotated development set, while execution, validation, stopping, and deployment are handled by deterministic components. This separation allows \system to use LLMs for discovering candidate repair mechanisms, without relying on them to directly clean the full table.

\noindent\textbf{Evidence construction.}
Stages (1)--(2) construct the evidence used by the synthesis loop. Given a raw table, \system first creates a compact annotated development set. The Dev-Set Sizer determines the annotation budget using table size, observed error coverage, and column-level diversity, after which human annotation produces trusted dirty-clean examples. The Dataset Profiler then summarizes the annotated table into compact evidence, including column statistics, missingness, frequent and rare values, type signatures, parseability under dates, times, and numbers, and candidate structural dependencies. Rather than passing the full table to the LLM agents, \system exposes this structured profile together with representative examples.

\noindent\textbf{Program synthesis loop.}
Stages (3)--(7) form an iterative synthesis and refinement loop. The Diagnose Agent identifies likely error types, affected columns, column dependencies, representative examples, and candidate repair opportunities. The Strategize Agent turns this diagnosis into an ordered cleaning strategy that specifies target columns, cleaning skills, guard predicates, and repair order. The Coding Agent compiles the strategy into a Python program that preserves the input schema, row order, and table shape, with each update protected by explicit guards. Candidate programs are executed in isolation. Syntax and runtime failures are repaired by a bounded debugging loop. For each executable candidate, the Rule-Based Decider evaluates cell-level metrics (e.g., precision, F1) on the development set. If the candidate is incomplete or unsafe, the Review Agent converts validation failures into targeted feedback for the next iteration.

\noindent\textbf{Deployment.}
In stage (8), \system commits the best validated cleaning program and applies it to the full raw table. This deployment path does not invoke any LLM. The committed program is inspectable, versionable, cacheable, and reusable on future schema-compatible batches. Thus, \system pays the LLM cost during synthesis, while repeated cleaning reduces to deterministic program execution.

The dashed box in Figure~\ref{fig:workflow} contains modules shared across datasets and iterations. Agent Memory is not shown as a separate numbered stage because it supports the entire synthesis loop, maintaining the current profile, diagnosis, strategy, generated program, validation feedback, and best validated program, and exposing role-specific context to each agent.

\section{Methodology}
\label{sec:methodology}


\subsection{FSM-based Cleaning Program Synthesis}
\label{sec:fsm}

Directly asking an LLM to clean a table conflates several
responsibilities, namely diagnosing data errors,
writing executable code, debugging failures, and deciding whether a
candidate repair is safe to deploy. \system instead formulates
program synthesis as a feedback-grounded finite-state machine
(FSM). LLM agents propose intermediate artifacts, while execution,
validation, and stopping decisions are grounded in deterministic
tools and development-set feedback.

Let $\Phi$ be the profile extracted from the annotated development set $\devset$, $\mathcal{K}$ be the cleaning-skills library, $\mathcal{B}$ be the role-aware memory buffer, and $\mathcal{X}$ be the isolated code execution environment. The FSM is defined as
\begin{equation}
\mathcal{M}_{\system}
=
(\mathcal{S},\mathcal{U},\mathcal{O},\delta,S_1,\mathsf{Acc}),
\end{equation}
where $\mathcal{S}=\{S_1,\ldots,S_6\}$ is the set of workflow
states, $\mathcal{U}$ is the set of agents and deterministic tools,
$\mathcal{O}$ is the set of transition actions, $\delta:
\mathcal{S}\times\mathcal{O}\rightarrow\mathcal{S}$ is the
transition function, $S_1$ is the initial state, and
$\mathsf{Acc}=\{S_6\}$ is the accepting state. The FSM operates in
the environment
\begin{equation}
\mathcal{E}_{\system}
=
(\dirty,\devset,\Phi,\mathcal{K},\mathcal{B},\mathcal{X}),
\end{equation}
which exposes table evidence, execution logs, and validation
metrics to the agents. At state $S_i$, the responsible agent emits an action
$O_k^{S_i}$, and the FSM transitions according to
\begin{equation}
\delta(S_i, O_k^{S_i}) \rightarrow S_j .
\end{equation}
The default forward transition of each state is denoted by
$O_0^{S_i}$. Non-default actions encode debugging, rejection, and
feedback-driven replanning.

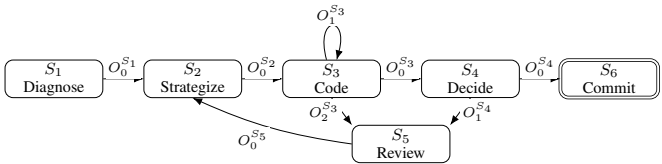
\begin{figure}[h]
\centering
\resizebox{\columnwidth}{!}{
\begin{tikzpicture}[
    >=Latex,
    node distance=7mm and 7mm,
    state/.style={
        draw,
        rounded corners,
        minimum width=17mm,
        minimum height=6.5mm,
        align=center,
        font=\footnotesize,
        inner sep=1.2pt
    },
    accept/.style={
        state,
        double,
        double distance=1pt
    },
    edge/.style={->,line width=0.35pt},
    lab/.style={font=\scriptsize,inner sep=1pt,fill=white}
]
\node[state] (s1) {$S_1$\\Diagnose};
\node[state, right=7mm of s1] (s2) {$S_2$\\Strategize};
\node[state, right=7mm of s2] (s3) {$S_3$\\Code};
\node[state, right=7mm of s3] (s4) {$S_4$\\Decide};
\node[accept, right=7mm of s4] (s6) {$S_6$\\Commit};
\node[state, below=8mm of $(s3)!0.5!(s4)$] (s5) {$S_5$\\Review};

\draw[edge] (s1) -- node[lab,above] {$O_0^{S_1}$} (s2);
\draw[edge] (s2) -- node[lab,above] {$O_0^{S_2}$} (s3);
\draw[edge] (s3) -- node[lab,above] {$O_0^{S_3}$} (s4);
\draw[edge] (s4) -- node[lab,above] {$O_0^{S_4}$} (s6);

\draw[edge] (s3) edge[loop above,min distance=8mm]
    node[lab] {$O_1^{S_3}$} (s3);
\draw[edge] (s3.south) -- node[lab,left] {$O_2^{S_3}$} (s5.north west);
\draw[edge] (s4.south) -- node[lab,right] {$O_1^{S_4}$} (s5.north east);

\draw[edge] (s5.west) to[bend left=8]
    node[lab,below left] {$O_0^{S_5}$} (s2.south);

\end{tikzpicture}
}
\caption{FSM workflow of \system. Edge label $O_k^{S_i}$ denotes
the $k$-th action of state $S_i$. $O_0^{S_i}$ is the default forward
action. $S_3$ has a bounded debugging self-loop. The double ring
marks $S_6$ as the accepting state.}
\label{fig:tabclean-fsm}
\end{figure}

\begin{table}[t]
\centering
\small
\caption{Key non-default transition conditions in the \system FSM.}
\label{tab:tabclean-key-transitions}
\begin{tabularx}{\columnwidth}{@{}l X@{}}
\toprule
Action & Condition \\
\midrule
$O_1^{S_3}$ &
The candidate program fails syntax or runtime checks and the
debugging budget is not exhausted. \\

$O_2^{S_3}$ &
The candidate program still fails after the debugging budget is
exhausted. \\

$O_1^{S_4}$ &
The candidate is unsafe, introduces regressions, or does not satisfy
the stopping condition. \\

$O_0^{S_4}$ &
The best validated program satisfies the target score, or the
iteration/patience budget is exhausted. \\
\bottomrule
\end{tabularx}
\end{table}

Figure~\ref{fig:tabclean-fsm} shows the FSM workflow, and Table~\ref{tab:tabclean-key-transitions} summarizes the conditions for each non-default FSM transition. The important distinction is that syntax and runtime failures are handled locally in $S_3$, while semantic failures are routed through $S_5$ so that the next strategy can change the repair clauses or their guards. When the FSM reaches $S_6$, the controller commits the best validated program $\prog^*$ and applies it to the raw table.
\begin{equation}
\widehat{\clean} = \prog^*(\dirty).
\end{equation}
Because $\prog^*$ is expressed as schema-level guarded repair logic, later schema-compatible tables can reuse it without additional LLM inference.

Algorithm~\ref{alg:tabclean-synthesis} summarizes the operational loop. \system does not trust an LLM-generated program directly. Each candidate must pass isolated
execution, cell-level validation, and safety checks. The controller maintains $\prog^*$ throughout synthesis. This best-program tracking is necessary because iterative LLM synthesis is not monotonic. A later candidate can improve recall on one column while introducing false positives or regressions on another.
```latex
\SetKwInput{KwIn}{Input} 
\SetKwInput{KwOut}{Output} 
\SetKw{KwBreak}{break} 
\SetKw{KwContinue}{continue} 
\SetKw{KwReturn}{return}

\begin{algorithm}[t]
\caption{\system Program Synthesis}
\label{alg:tabclean-synthesis}

\KwIn{Dirty table $\dirty$, development set $\devset$, skill library $\mathcal{K}$,
iteration budget $T$, debugging budget $B$}
\KwOut{Reusable cleaning program $\prog^*$ and cleaned table $\widehat{\clean}$}

$\Phi \leftarrow \mathrm{Profile}(\devset)$\;
$d \leftarrow \mathrm{Diagnose}(\devset,\Phi,\mathcal{K})$\;
$r \leftarrow \mathrm{Strategize}(d,\mathcal{K})$\;
$\prog^* \leftarrow \bot$; $s^* \leftarrow 0$\;

\BlankLine
\For{$t \leftarrow 1$ \KwTo $T$}{
    $\prog_t \leftarrow \mathrm{Code}(r,\prog^*)$\;
    $ok \leftarrow \mathrm{false}$\;

    \For{$b \leftarrow 1$ \KwTo $B$}{
        $(ok, log, out) \leftarrow \mathrm{Execute}(\prog_t,\devset)$\;
        \If{$ok$}{
            \KwBreak\;
        }
        $\prog_t \leftarrow \mathrm{Debug}(\prog_t,log)$\;
    }

    \If{$\neg ok$}{
        $q \leftarrow \mathrm{ReviewFailure}(\prog_t,log)$\;
        $r \leftarrow \mathrm{Strategize}(d,q,\mathcal{K})$\;
        \KwContinue\;
    }

    $(s_t,F_t) \leftarrow \mathrm{Evaluate}(out,\devset)$\;

    \If{$s_t > s^*$ \textnormal{\bfseries and} $\mathrm{Safe}(F_t)$}{
        $\prog^* \leftarrow \prog_t$\;
        $s^* \leftarrow s_t$\;
    }

    \If{$\mathrm{Stop}(s^*,t,F_t)$}{
        \KwBreak\;
    }

    $q \leftarrow \mathrm{Review}(\prog_t,F_t,\prog^*)$\;
    $r \leftarrow \mathrm{Strategize}(d,q,\mathcal{K})$\;
}

\BlankLine
$\widehat{\clean} \leftarrow \prog^*(\dirty)$\;
\KwReturn{$\prog^*, \widehat{\clean}$}\;
\end{algorithm}

\subsection{Repair Program Representation and Semantics}
\label{sec:program-semantics}

The central artifact in \system is an executable repair program, not a list of predicted repaired cells. A program is an ordered sequence of guarded clauses.
\begin{equation}
\prog = \langle c_1,\ldots,c_k\rangle,
\qquad
c_\ell = (A_\ell, s_\ell, \guard_\ell, f_\ell, e_\ell, \pi_\ell).
\end{equation}
Here $A_\ell$ is the target column, $s_\ell$ is a repair skill selected from the library $\mathcal{K}$, $\guard_\ell$ is a predicate that determines whether the clause may fire, $f_\ell$ is a deterministic transformation, $e_\ell$ records the supporting evidence, and $\pi_\ell$ specifies the clause priority or application order. The Strategize Agent produces this clause-level plan, and the Coding Agent compiles it into an executable Python script.

For a row $x_i$ and target column $A_\ell$, a clause has the semantics
\begin{equation}
c_\ell(x_i)_{A_\ell}=
\begin{cases}
f_\ell(x_i,A_\ell), & \text{if } \guard_\ell(x_i,A_\ell,\Phi,e_\ell)=1,\\
x_{iA_\ell}, & \text{otherwise.}
\end{cases}
\end{equation}
All non-target columns are left unchanged by the clause. A guard may depend on the cell value, other columns in the same row, column-level profile statistics, and evidence extracted from the development set. It must not depend on row identifiers, held-out labels, or manually enumerated test repairs. The program applies clauses in strategy order and preserves the input schema, row order, and table shape.
\begin{equation}
\mathrm{schema}(\prog(\dirty))=\mathrm{schema}(\dirty),
\qquad
|\prog(\dirty)|=|\dirty|.
\end{equation}

This representation differs from direct value editing in two ways. First, the unit of synthesis is a general repair mechanism, such as time normalization or dependency-based majority repair, rather than an individual repaired cell. Second, every transformation is paired with an explicit guard. The guard is the main mechanism for precision: it requires the program to establish that a value matches a supported dirty pattern before changing it. A case study is provided in Section~\ref{sec:program-case-study}.

\subsection{Evidence Construction}
\label{sec:evidence}

\system constructs evidence before invoking the main synthesis loop. This evidence serves two purposes. It reduces the prompt context that agents must inspect, and it constrains repair proposals to mechanisms supported by the table.

\subsubsection{Adaptive Development-Set Sizer}

The development set should expose recurring errors without becoming a second full-labeling task. \system uses an adaptive sizer that samples rows until several coverage conditions are met, namely a minimum number of rows, a minimum number of dirty cells, coverage of error-bearing columns, and sufficient examples for high-frequency error types. A hard cap prevents the development set from becoming a large fraction of the full table.

The sizer also records undercovered columns. A column is undercovered when the current development set contains too few labeled dirty cells to support reliable diagnosis and validation for that column. Downstream agents treat such columns conservatively. They may still propose guards for deterministic format errors, but they avoid broad dependency or semantic repairs unless additional evidence is available. This design makes insufficient annotation primarily reduce recall rather than precision.

\subsubsection{Dataset Profiler}

The profiler computes compact statistics over the development set. For each column, it records cardinality, missingness, frequent and rare values, type signatures, string-length distributions, numeric ranges, and parseability as dates, times, numbers, or units. It also searches for repeated identifiers, nearly functional dependencies, group-level constants, common composite-field structures, and column pairs that may support evidence-based repair.

The profile is not a dump of the table. It contains summaries and representative examples selected to support diagnosis. For example, if a city column almost determines a state column, and the development labels show that minority state values are dirty, the profile records a candidate dependency repair. If several values are parseable as times and the clean examples share a target format, the profile records a format-normalization opportunity.

\subsubsection{Evidence Levels}

The Diagnose Agent classifies repair opportunities by evidence level.
\begin{itemize}
    \item \textbf{Deterministic evidence}. The dirty pattern and clean target are mechanically inferable, such as date, time, numeric, unit, or whitespace normalization with an unambiguous target format.
    \item \textbf{Table-backed evidence}. The repair follows from cross-row or cross-column regularities, such as majority values within an entity group or near-functional dependencies.
    \item \textbf{Weak or external evidence}. The correct value cannot be inferred from the development set and table profile alone, such as sparse address repairs or ambiguous entity names.
\end{itemize}
Only the first two evidence levels are eligible for automatic program synthesis. Weak or external repairs are skipped or marked for future human or retrieval-backed validation. This classification prevents the LLM from turning plausible world knowledge into unsupported table edits.

\subsection{Skill-Guided Guard Synthesis}
\label{sec:guard-synthesis}

The cleaning-skills library $\mathcal{K}$ encodes reusable cleaning knowledge as repair contracts. A skill is not executed directly. Instead, it constrains the diagnosis, strategy, coding, and review stages by specifying what evidence is required, what guard must be constructed, and what transformation is allowed. Table~\ref{tab:skill-contracts} shows representative skills.

\begin{table*}[t]
\centering
\scriptsize
\caption{Representative cleaning-skill contracts used by \system. A skill constrains repair synthesis by requiring evidence and guards before code can be generated.}
\label{tab:skill-contracts}
\begin{tabularx}{\textwidth}{@{}l l X X X@{}}
\toprule
\textbf{Skill} & \textbf{Error type} & \textbf{Required evidence} & \textbf{Guard} & \textbf{Transformation} \\
\midrule
\textsc{FormatNormalize} &
FI &
Dirty and clean examples reveal a target representation, such as date, time, or numeric format. &
Value is parseable under supported formats and does not already match the target format. &
Parse and emit canonical string. \\

\textsc{UnitScaleRepair} &
UT &
Numeric values show a consistent scale or unit mismatch with clean examples. &
Value falls in the suspicious range or has an explicit unit marker. &
Convert to target unit or scale. \\

\textsc{MajorityFDRepair} &
VAD &
A key or column group nearly determines the target column in the development profile. &
Group majority exists, support exceeds threshold, and current value disagrees. &
Replace by supported group majority. \\

\textsc{AliasCanonicalize} &
SE &
Aliases are supported by repeated examples or profile-level canonicalization evidence. &
Value belongs to a validated alias set and target canonical form is unique. &
Map alias to canonical value. \\

\textsc{MissingFill} &
MV &
A deterministic source exists in another column, group, or profile-backed dependency. &
Target value is missing and source value is available and unambiguous. &
Fill missing value from source. \\

\textsc{CompositeSplitMerge} &
CM &
Examples show repeated composite-field structure or extraction artifact. &
Value matches the composite pattern and target fields are empty or inconsistent. &
Extract, split, merge, or realign fields. \\
\bottomrule
\end{tabularx}
\end{table*}

A guarded repair usually combines several guard families.
\begin{equation}
\Guard
=
\guard_{\mathrm{dirty}}
\wedge
\guard_{\mathrm{target}}
\wedge
\guard_{\mathrm{evidence}}
\wedge
\guard_{\mathrm{scope}}.
\end{equation}
The dirty-pattern guard $\guard_{\mathrm{dirty}}$ checks whether the value matches a supported error pattern. The target-negative guard $\guard_{\mathrm{target}}$ prevents already-clean cells from being modified. The evidence guard $\guard_{\mathrm{evidence}}$ requires sufficient support from development labels or profile statistics. The scope guard $\guard_{\mathrm{scope}}$ restricts the repair to the intended column, group, or value range. Table~\ref{tab:guard-families} summarizes these guard families.

\begin{table}[t]
\centering
\small
\caption{Guard families used to prevent over-correction.}
\label{tab:guard-families}
\begin{tabularx}{\columnwidth}{@{}l X@{}}
\toprule
\textbf{Guard family} & \textbf{Purpose} \\
\midrule
Dirty-pattern guard &
Requires the value to match an observed or profile-supported dirty pattern. \\

Target-negative guard &
Skips values that already satisfy the target canonical representation. \\

Evidence guard &
Requires sufficient development-set or profile support before applying a repair. \\

Scope guard &
Restricts the repair to the intended column, key group, type range, or parseable domain. \\

Anti-regression guard &
Added after validation feedback to block transformations that produced false positives. \\
\bottomrule
\end{tabularx}
\end{table}

The first strategy is intentionally conservative. It prioritizes deterministic repairs because they can be validated with few examples and typically admit precise guards. Later strategies expand coverage only when validation feedback identifies persistent false negatives and the review memo explains them as missed supported mechanisms. Conversely, when a candidate introduces false positives, the next strategy narrows or removes the offending clause, often by adding anti-regression guards.

\subsection{Constrained Code Synthesis}
\label{sec:constrained-code}

The Coding Agent compiles a strategy into a complete Python program. Unlike free-form code generation, \system imposes a program contract.
\begin{itemize}
    \item The program must read the input table as strings and avoid unintended type coercion.
    \item The output must preserve the original schema, row order, and table shape.
    \item Each cell update must be protected by an explicit guard.
    \item Transformations must be expressed through column values, patterns, or structural relationships.
    \item The program must not hard-code row identifiers, held-out values, or development-set lookup tables.
    \item Unsupported or ambiguous values must be left unchanged.
\end{itemize}

When a best program $\prog^*$ already exists, the Coding Agent edits incrementally rather than generating from scratch. The review memo marks clauses as \textsc{Keep}, \textsc{Add}, \textsc{Modify}, \textsc{Remove}, or \textsc{Accept}. Logic marked \textsc{Keep} should be copied without modification because it has already passed validation. Logic marked \textsc{Modify} should be narrowed or corrected according to FP/FN feedback. This incremental editing protocol reduces oscillation and makes each iteration attributable. If performance changes, the controller can relate the change to a small number of modified clauses.

Syntax and runtime failures are handled separately from semantic cleaning errors. The Debugger receives the current script and execution log, then returns a patched script. It may fix imports, parsing edge cases, missing columns, type errors, and file I/O issues. It is not allowed to invent a new cleaning strategy unless the failure shows that the current strategy cannot be implemented under the program contract. This separation prevents the debugging loop from silently changing semantic repair intent.

\subsection{Execution-Guided Validation and Control}
\label{sec:validation-control}

Candidate programs are executed before they are judged. The Executor materializes the candidate as a temporary script, binds \texttt{INPUT\_PATH} and \texttt{OUTPUT\_PATH}, runs it with a timeout, and checks that the produced table has the same columns, row order, and shape as the input. Failed executions are routed to the Debugger. Successful executions are compared against the labeled development set at cell level.

For iteration $t$, the evaluator returns
\begin{equation}
F_t=(TP_t,FP_t,FN_t,\Delta_t,M_t),
\end{equation}
where $TP_t$, $FP_t$, and $FN_t$ are cell-level repair counts of true positives, false positives, and false negatives. $\Delta_t$ summarizes regressions relative to the best program, and $M_t$ stores per-column failure messages. Development-set precision, recall, and F1 are computed as $P_t=TP_t/(TP_t+FP_t)$, $Q_t=TP_t/(TP_t+FN_t)$, and $F1_t=2P_tQ_t/(P_t+Q_t)$.

Among executable programs, the controller tracks the best development-set score.
\begin{equation}
\prog^*
=
\arg\max_{\prog_t:\mathrm{Safe}(F_t)=1} F1_t.
\end{equation}
The stopping rule fires when the best score reaches the target threshold, the iteration budget is exhausted, or no improvement has been observed for a patience window. Otherwise, the candidate, feedback, and best-program summary are passed to the Review Agent.

The review memo is structured rather than conversational. For false negatives, it distinguishes missed mechanisms from implementation bugs and cases with insufficient evidence. For false positives, it identifies over-broad transformations and missing guards. For regressions, it points to the clause that should be restored from $\prog^*$. The memo ends with a column-level action table using labels such as \textsc{Keep}, \textsc{Modify}, \textsc{Remove}, or \textsc{Accept}. This makes the next strategy update grounded in executed behavior rather than in the LLM's self-assessed confidence.

\subsection{Role-Specialized Agents and Memory}
\label{sec:agents-memory}

\system uses multiple role-specialized agents because different stages require different context and have different failure modes. This design follows the broader pattern of tool-grounded LLM systems, where LLMs propose plans or programs while deterministic tools perform execution and validation~\cite{cleanagent,cocoon,quite,sweagent,selfdebugging,reflexion}. In \system, agents do not share an unfiltered transcript. Instead, the FSM maintains a role-aware memory buffer $\mathcal{B}$ containing the current profile, diagnosis, strategy, candidate program, best program, execution feedback, and review history.

\noindent\textbf{Diagnose Agent.}
The Diagnose Agent receives the profile, development-set errors, error taxonomy, and skills library. It outputs a structured diagnosis containing column-level error types, representative examples, candidate dependencies, evidence levels, and repair opportunities. It also marks unsupported columns or mechanisms whose correct repairs would require external knowledge.

\noindent\textbf{Strategize Agent.}
The Strategize Agent turns the diagnosis into an executable repair plan. The plan specifies target columns, selected skills, guards, application order, and skip decisions. In later iterations, the agent receives the review memo and best prior strategy, then performs an incremental update rather than replanning from scratch. This preserves validated repairs and limits the number of simultaneous changes.

\noindent\textbf{Coding Agent and Debugger.}
The Coding Agent translates the plan into Python under the program contract in Section~\ref{sec:constrained-code}. The Debugger is invoked only after syntax or runtime failure. It patches the current script using execution logs while preserving the high-level strategy.

\noindent\textbf{Rule-Based Decider.}
The Decider is deterministic. It computes development-set metrics, applies safety checks, updates $\prog^*$ when appropriate, and determines whether to stop or continue. This component prevents a fluent but unsafe LLM-generated program from being committed.

\noindent\textbf{Review Agent.}
The Review Agent closes the semantic feedback loop. It receives the executed program, cell-level FP/FN summaries, regression information, and the current best program. Its output explains why the candidate failed or improved, then converts that explanation into actionable edits for the next strategy.

\noindent\textbf{Agent Memory}
Passing the full interaction trace to every agent is noisy and expensive \cite{longcontext,babilong,quite}. Outdated code, verbose logs, and stale reasoning distract agents as prompts grow across iterations. \system\ instead retains only high-value context. The FSM state serves as
a role-aware working memory that stores the current profile,
diagnosis, strategy, generated program, execution feedback,
evaluation history, and the best validated program so far. Each agent receives only the context needed for its role. The
Diagnose Agent observes the profile, sampled errors, structural
signals, and cleaning skills. The Strategize Agent receives the
diagnosis, recent feedback, evaluation summary, and best prior
strategy. The Coding Agent receives the current strategy, relevant
skills, profile, best code, and recent implementation feedback. The
Review Agent receives the executed code, cell-level FP/FN feedback,
diff summary, and iteration dashboard. This role-scoped memory reduces
prompt cost while letting later iterations preserve successful repairs
and avoid repeating failed ones.

\section{Experimental Evaluation}
\label{sec:experiments}

Our evaluation answers six research questions. \textbf{RQ1} compares \system with rule-based, learning-based, and LLM-based baselines in precision, recall, and F1. \textbf{RQ2} measures whether compiling repairs into code reduces end-to-end runtime on small and large tables. \textbf{RQ3} analyzes token usage and API cost relative to LLM-based baselines. \textbf{RQ4} ablates the finite-state multi-agent workflow, role-aware memory, and best-program tracking. \textbf{RQ5} studies the sensitivity to development-set size. \textbf{RQ6} evaluates the effect of model choice on quality and convergence.

\subsection{Experimental Setup}
\label{sec:setup}

\noindent\textbf{Datasets.} Table~\ref{tab:dataset_statistics} summarizes the six benchmarks, spanning healthcare, aviation, beverages, tax records, literature screening, and movie metadata. They cover both small and large tables and include heterogeneous error patterns, ranging from typos and format inconsistencies to functional-dependency violations and missing values. The number of development-set rows is set by our adaptive sizer.

\begin{table}[t]
\centering
\small
\caption{Datasets used in the experiments.}
\label{tab:dataset_statistics}
\begin{tabular}{lrrrr}
\toprule
\textbf{Dataset} & \textbf{Rows} & \textbf{Cols.} & \textbf{Err. cells} & \textbf{Dev rows} \\
\midrule
hospital~\cite{holistic,hospitalcompare} & 1,000 & 19 & 509 & 150\\
flight~\cite{flight} & 2,376 & 7 & 9,504 & 110 \\
beers~\cite{craftcans_kaggle} & 2,410 & 10 & 4,765 & 240 \\
tax~\cite{tax,bart} & 200,000 & 15 & 121,219 & 150 \\
rayyan~\cite{rayyan} & 1,000 & 11 & 960 & 150\\
movies~\cite{movies} & 7,390 & 17 & 7,675 & 369 \\
\bottomrule
\end{tabular}
\par\smallskip
\begin{minipage}{\columnwidth}
\footnotesize ``Err.\ cells'' is the number of erroneous cells in the \emph{full} table. Cleaning quality is evaluated on held-out rows, so the $TP{+}FN$ reported in later tables is slightly smaller. The difference corresponds to errors that fall in the development set.
\end{minipage}
\end{table}

\noindent\textbf{Baselines.} We compare against four representative systems. HoloClean is a probabilistic data repairing system that combines constraints and statistical signals \cite{holoclean}. Baran is a semi-supervised correction engine that learns repair transformations from data and labeled examples \cite{baran}. Cocoon uses an LLM to generate executable cleaning logic and validate it in a feedback loop \cite{cocoon}. IterClean is an iterative LLM-based pipeline that alternates between error detection, verification, and repair \cite{iterclean}. These baselines cover rule-based/probabilistic, learning-based, and LLM-based paradigms. We used the code repositories released with the baseline papers and followed their provided execution instructions.

\noindent\textbf{Human input.} Table~\ref{tab:human-labor} summarizes the human input required by each system per dataset. \system uses more up-front labels than seed-based LLM cleaners, but those labels are used to synthesize a reusable program rather than to clean only one table pass.

\setlength{\textfloatsep}{8pt plus 2pt minus 2pt}
\setlength{\floatsep}{6pt plus 2pt minus 2pt}
\setlength{\intextsep}{6pt plus 2pt minus 2pt}
\setlength{\abovecaptionskip}{2pt}
\setlength{\belowcaptionskip}{0pt}
\begin{table}[h]
\centering
\small
\setlength{\tabcolsep}{4pt}
\caption{Required human input per dataset.}
\resizebox{\columnwidth}{!}{%
\begin{tabular}{l l c}
\toprule
\textbf{System} & \textbf{Required Human Input} & \textbf{Avg. Time} \\
\midrule
HoloClean  & Denial constraints / FDs ($\sim$5--10 rules) & 1--2 h \\
Baran     & Cell labels on 20 tuples (active learning)  & 5--10 min \\
Cocoon    & None                                        & 0 \\
IterClean & Cell labels on 5 seed tuples                & 2--5 min \\
\system   & Cell labels on adaptive dev set     & 1--2 h \\
\bottomrule
\end{tabular}%
}

\label{tab:human-labor}
\end{table}

\noindent\textbf{Metrics.} Cleaning quality is measured by cell-level precision, recall, and F1 on held-out rows. A dirty cell repaired to an incorrect value remains a false negative, while only changes to originally clean cells count as false positives. Runtime measures end-to-end wall-clock time, including profiling, LLM calls, code generation, execution, debugging, and evaluation. For LLM-based methods, we report input tokens, output tokens, and API cost when available. 

\noindent\textbf{Implementation.} We implement \system as a Python prototype for CSV-based tabular cleaning. The pipeline uses \texttt{pandas} for table I/O, profiling, and evaluation, and a LangGraph controller to execute the FSM-based workflow. Input values are read as strings to avoid unintended type conversion, and generated programs are required to preserve the original schema, row order, and table shape. 

\noindent \textbf{Model Selection.} We employ gpt-5-mini for both Cocoon and IterClean, as well as for the diagnose and review agents in \system. We use gpt-5.3-codex for the coding agent and gpt-5.2 for the strategize agent.

\subsection{RQ1: Cleaning Effectiveness}
\label{sec:effectiveness}

\begin{table*}[t]
\centering
\scriptsize
\setlength{\tabcolsep}{3.5pt}
\caption{Data cleaning performance (P = precision, R = recall, F = F1, higher is better). \textbf{Bold} marks the best F1 per dataset (ties included). All scores are rounded to two decimals, so some small nonzero values are displayed as $0.00$.}
\resizebox{\textwidth}{!}{%
\begin{tabular}{l ccc ccc ccc ccc ccc ccc}
\toprule
\textbf{System}
& \multicolumn{3}{c}{\textbf{hospital}}
& \multicolumn{3}{c}{\textbf{flight}}
& \multicolumn{3}{c}{\textbf{beers}}
& \multicolumn{3}{c}{\textbf{tax}}
& \multicolumn{3}{c}{\textbf{rayyan}}
& \multicolumn{3}{c}{\textbf{movies}} \\
\cmidrule(lr){2-4}\cmidrule(lr){5-7}\cmidrule(lr){8-10}\cmidrule(lr){11-13}\cmidrule(lr){14-16}\cmidrule(lr){17-19}
& P & R & F
& P & R & F
& P & R & F
& P & R & F
& P & R & F
& P & R & F \\
\midrule
HoloClean
& 1.00 & 0.91 & \textbf{0.95}
& 0.00 & 0.00 & 0.00
& 1.00 & 0.04 & 0.07
& 0.82 & 0.01 & 0.01
& 0.00 & 0.00 & 0.00
& 0.01 & 0.00 & 0.00 \\
Baran
& 1.00 & 0.54 & 0.70
& 1.00 & 0.19 & 0.32
& 1.00 & 0.78 & 0.88
& 1.00 & 0.73 & 0.84
& 1.00 & 0.22 & 0.36
& 1.00 & 0.67 & 0.80 \\
Cocoon
& 1.00 & 0.42 & 0.59
& 0.00 & 0.00 & 0.00
& 1.00 & 0.03 & 0.07
& 0.00 & 0.00 & 0.00
& 0.00 & 0.00 & 0.00
& 0.00 & 0.00 & 0.00 \\
IterClean
& 1.00 & 0.91 & \textbf{0.95}
& 0.00 & 0.00 & 0.00
& 1.00 & 0.60 & 0.75
& 0.03 & 0.01 & 0.02
& 0.01 & 0.00 & 0.00
& 0.21 & 0.05 & 0.08\\
\textbf{\system}
& 1.00 & 0.77 & 0.87
& 1.00 & 0.68 & \textbf{0.81}
& 1.00 & 0.89 & \textbf{0.94}
& 1.00 & 0.98 & \textbf{0.99}
& 1.00 & 0.85 & \textbf{0.92}
& 0.99 & 0.83 & \textbf{0.91}\\
\bottomrule
\end{tabular}%
}

\label{tab:our-baseline-prf}
\end{table*}

Table~\ref{tab:our-baseline-prf} shows that \system attains the best F1 on five of six benchmarks and remains competitive on \textit{hospital}. It keeps precision at 1.00 on every dataset except \textit{movies}, where precision is 0.99, while improving recall on large and heterogeneous tables. The contrast is most visible on \textit{tax}: Cocoon rounds to 0.00 F1, IterClean reaches 0.02 F1 under the sample-scaled tax evaluation, while \system reaches 0.99 F1 on the full table. On \textit{rayyan} and \textit{movies}, IterClean reaches 0.00 and 0.08 F1, whereas \system reaches 0.92 and 0.91. Values printed as 0.00 may be exact zeros or small nonzero scores rounded to two decimals.

HoloClean performs well on \textit{hospital} but has zero or near-zero recall elsewhere, even with dataset-specific denial constraints. The main limitation is candidate reachability. HoloClean can select from generated candidate domains, but many benchmarks require normalized strings that are absent from those domains. In our runs, the correct value appeared in the candidate domain for 98.8\% of \textit{hospital} error cells but only 0.0\% of \textit{flight}, 19.6\% of \textit{beers}, 1.6\% of \textit{tax}, 2.0\% of \textit{rayyan}, and 11.4\% of \textit{movies}. Cocoon's regular-expression and SQL \texttt{CASE WHEN} rules cover only a small fraction of heterogeneous errors, while IterClean's per-cell prompting is expensive at scale and over-edits or mislabels free-form bibliographic and metadata fields.

The largest gains appear on \textit{rayyan}, \textit{flight}, and \textit{tax}. \textit{movies} also improves over Baran and substantially outperforms IterClean. \textit{flight} contains systematic time and dependency errors, and \textit{tax} has regular formatting and canonicalization errors, both of which are well suited to synthesized programs. \textit{rayyan} and \textit{movies} are harder because they include free-form fields, but \system recovers high recall through guarded structural repairs and cautious normalization. On \textit{hospital}, \system is below the strongest baselines, suggesting that sparse identifier and address-like fields require stronger external evidence than \system currently uses.

\subsection{RQ2: Runtime Efficiency}
\label{sec:runtime}

\begin{figure*}[t]
\centering
\includegraphics[width=\textwidth]{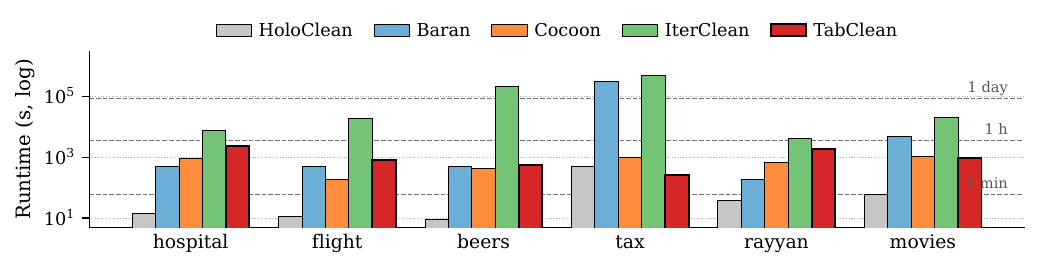}
\caption{End-to-end wall-clock runtime (seconds, log scale, lower is better). Dashed lines mark the 1\,min, 1\,h, and 1\,day reference levels.}
\label{tab:our-baseline-time}
\end{figure*}

Figure~\ref{tab:our-baseline-time} shows that \system remains practical despite using multiple agents. The reason is that LLM calls occur only during diagnosis, planning, code generation, debugging, and review. Once a candidate program exists, applying it to thousands or hundreds of thousands of rows is ordinary Python execution. Compared with Baran, \system reduces \textit{tax} runtime from 89.8 hours to 4.42 minutes while improving F1 from 0.84 to 0.99. Against IterClean on the 200,000-row \textit{tax} table, \system is about 1,913$\times$ faster. Compared with IterClean, \system also reduces \textit{beers} runtime from 58.3 hours to 9.32 minutes while improving F1 from 0.75 to 0.94.

The updated IterClean runs on \textit{rayyan} and \textit{movies} further support this trend. \system finishes \textit{rayyan} in 32.4 minutes compared with 71.0 minutes for IterClean, a 2.19$\times$ speedup, while also raising F1 from 0.00 to 0.92. On \textit{movies}, \system finishes in 16.0 minutes compared with 5.92 hours for IterClean, a 22.2$\times$ speedup, and is also faster than Baran and Cocoon while achieving higher F1. HoloClean is faster on several small datasets but has much lower recall, so its runtime is not directly comparable at similar quality. Cocoon is competitive on small tables, but its repairs often fail to generalize beyond narrow patterns. The runtime profile of \system depends primarily on iteration count and code-generation cost rather than table size. This is visible on \textit{tax}, where the error mechanism is regular and the system terminates quickly even though the table is large.

\subsection{RQ3: Token Usage and API Cost}
\label{sec:cost}

\begin{figure}[t]
\centering
\includegraphics[width=\columnwidth]{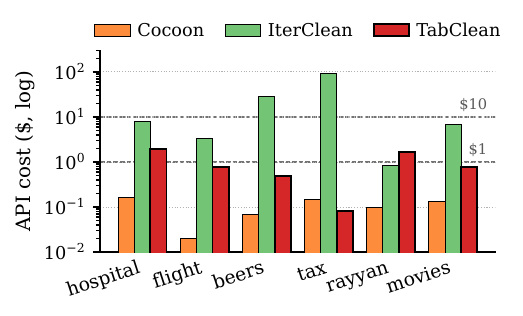}
\caption{API cost for the LLM-based systems (log scale, lower is better). \system\ stays below two dollars on every dataset, while IterClean's per-table cost grows into the tens of dollars on \textit{beers} and \textit{tax}.}
\label{fig:cost}
\end{figure}

\begin{table}[t]
\centering
\caption{LLM token usage (thousands of tokens) for the LLM-based systems. Per-dataset API cost is shown in Figure~\ref{fig:cost}.}
\label{tab:token-usage}
\begin{tabular}{l rr rr rr}
\toprule
& \multicolumn{2}{c}{\textbf{Cocoon}}
& \multicolumn{2}{c}{\textbf{IterClean}}
& \multicolumn{2}{c}{\textbf{\system}} \\
\cmidrule(lr){2-3}\cmidrule(lr){4-5}\cmidrule(lr){6-7}
\textbf{Dataset} & In & Out & In & Out & In & Out \\
\midrule
hospital & 14 & 78 & 3{,}727 & 3{,}571 & 520 & 143 \\
flight   & 5  & 14 & 1{,}303 & 1{,}553 & 337 & 53 \\
beers    & 8  & 32 & 7{,}560 & 13{,}096 & 196 & 37 \\
tax      & 20 & 72 & 87{,}251 & 36{,}643 & 25 & 8 \\
rayyan   & 12 & 50 & 1{,}008 & 303 & 599 & 138 \\
movies   & 21 & 65 & 13{,}807 & 1{,}633 & 298 & 66 \\
\bottomrule
\end{tabular}
\end{table}

Figure~\ref{fig:cost} shows the cost-quality advantage of compiling LLM reasoning into code, and Table~\ref{tab:token-usage} reports the underlying token usage. \system stays below two dollars on every dataset and costs eight cents on \textit{tax}, where IterClean spends \$93.49 to process the table once and Cocoon's \$0.15 run yields an F1 that rounds to 0.00. On \textit{tax}, \system therefore uses 0.086\% of IterClean's API cost. It is also cheaper on \textit{hospital}, \textit{flight}, \textit{beers}, and \textit{movies} (24.5\%, 23.0\%, 1.75\%, and 11.8\% of IterClean's cost). The exception is \textit{rayyan}, where \system costs \$1.66 against IterClean's \$0.85, but this buys an F1 improvement from 0.00 to 0.92. The gap should widen on larger tables and recurring batches: direct LLM cleaners pay per table pass, whereas \system pays once to synthesize a reusable program and then runs it without further LLM calls.

Cost also decomposes by agent role. Diagnosis and review are relatively inexpensive because they operate on compact summaries. Strategy and code generation are more expensive per call but occur only a few times. In the ablation study, we further measure whether lower-cost models can replace high-capability models without hurting quality.

\noindent\textbf{Amortized cost per repair.}
Figure~\ref{fig:cost} reports the one-time synthesis cost, not how it amortizes after deployment. We therefore report the API cost per 100 true-positive repairs, $\mathrm{Cost}_{100}=100\,C_{\mathrm{synth}}/TP$, where \(C_{\mathrm{synth}}\) is the total LLM cost before program commit and \(TP\) is the number of correctly repaired cells. This metric charges \system for all synthesis-time LLM calls but credits only correct repairs. On \textit{tax}, \(C_{\mathrm{synth}}=\$0.08\) and the committed program yields \(TP=118{,}150\) correct repairs, so $\mathrm{Cost}_{100}\approx\$6.8\times10^{-5}$, about 100 correct repairs per \(10^{-4}\) dollars of LLM cost. Even granting IterClean a perfect \(TP=118{,}150\), its \$93.49 \textit{tax} run gives $\mathrm{Cost}_{100}\approx\$0.079$, over a thousand times higher, and IterClean pays this on every table pass rather than once. Our estimate is in fact conservative, since it amortizes synthesis over a single table while the same program reruns on schema-compatible batches at no further LLM cost.

\subsection{RQ4: Ablation Study}
\label{sec:ablation}

\begin{table}[b]
\centering
\small
\setlength{\tabcolsep}{2pt}
\caption{RQ4 ablation results. Values are held-out F1. The w/o FSM variant replaces the finite-state multi-agent workflow with a single full-context code-generation prompt, w/o mem. removes role-aware memory management, and w/o best disables best-program tracking.}
\label{tab:ablation}
\begin{tabular}{lrrrr}
\toprule
\textbf{Dataset} & \textbf{\system} & \makecell{\textbf{w/o}\\\textbf{FSM}} & \makecell{\textbf{w/o}\\\textbf{memory}} & \makecell{\textbf{w/o best}\\\textbf{tracking}} \\
\midrule
hospital & 0.87 & 0.34 & 0.24 & 0.90 \\
flight & 0.81 & 0.66 & 0.68 & 0.68 \\
beers & 0.94 & 0.92 & 0.94 & 0.94 \\
tax & 0.99 & 0.99 & 0.84 & 0.84 \\
rayyan & 0.92 & 0.84 & 0.77 & 0.88 \\
movies & 0.91 & 0.89 & 0.90 & 0.87 \\
\midrule
\textbf{Avg.} & 0.91 & 0.77 & 0.73 & 0.85 \\
\bottomrule
\end{tabular}
\end{table}

To isolate the effect of each component, we evaluate variants of \system that remove or simplify one design choice at a time. One collapses the finite-state multi-agent workflow into a single full-context code-generation prompt (w/o FSM), one replaces role-aware memory with unfiltered conversation history, and one uses the latest program rather than the best validated program.

Table~\ref{tab:ablation} shows that collapsing \system into a single full-context code-generation prompt (w/o FSM) is not sufficient to match the full workflow. Its average F1 is 0.77, compared with 0.91 for \system. This variant remains strong on regular transformations such as \textit{tax}, \textit{beers}, and \textit{movies}, but drops sharply on \textit{hospital} and \textit{flight}, where repair requires coordinating heterogeneous evidence, guarded transformations, and validation feedback. This indicates that the benefit of \system is not merely from exposing the model to the table profile and development examples, but from decomposing the task into specialized roles and refining their output through validated synthesis stages.

Removing role-aware memory also causes a large degradation. Average F1 drops from 0.91 to 0.73, with especially large losses on \textit{hospital}, \textit{flight}, \textit{tax}, and \textit{rayyan}. These datasets require the agents to preserve earlier evidence about column dependencies, repair constraints, and prior failed attempts. Passing unfiltered history makes later strategy and code-generation steps more susceptible to stale or irrelevant context, so the system misses many repair opportunities while still keeping high precision.

Best-program tracking is the second most important stabilizer. Without it, average F1 falls to 0.85, mainly because late iterations can overwrite earlier high-recall programs on \textit{flight}, \textit{tax}, \textit{rayyan}, and \textit{movies}. This confirms that iterative LLM synthesis is not monotonic. Later programs may look plausible but regress on held-out repairs. Keeping the best development-set program separates exploration from deployment and prevents these regressions from becoming the final output.

\subsection{RQ5: Development-Set Sensitivity}
\label{sec:devsize}

\begin{table}[b]
\centering
\small
\setlength{\tabcolsep}{3pt}
\caption{RQ5 development-set sensitivity. Values are held-out F1, higher is better.}
\label{tab:devsize}
\begin{tabular}{lrrrr}
\toprule
\textbf{Dataset} & \textbf{$0.5\times$ adaptive} & \textbf{Adaptive} & \textbf{$2\times$ adaptive} & \textbf{Random 1\%} \\
\midrule
hospital & 0.61 & 0.87 & 0.85 & 0.13 \\
flight & 0.82 & 0.81 & 0.83 & 0.81 \\
beers & 0.92 & 0.94 & 0.94 & 0.90 \\
tax & 0.99 & 0.99 & 0.99 & 1.00 \\
rayyan & 0.94 & 0.92 & 0.95 & 0.86 \\
movies & 0.91 & 0.91 & 0.92 & 0.91 \\
\midrule
\textbf{Avg.} & 0.87 & 0.91 & 0.91 & 0.77 \\
\bottomrule
\end{tabular}
\end{table}

\system uses a small annotated development set, so we evaluate how quality changes with annotation budget. Table~\ref{tab:devsize} compares the adaptive sizer with three alternatives: half of the adaptive rows, twice the adaptive rows, and a random 1\% sample.

The adaptive sizer achieves a strong balance between annotation cost and cleaning quality. It reaches 0.91 average F1, while doubling the development set yields only a marginal gain. The extra labels slightly help \textit{flight}, \textit{rayyan}, and \textit{movies}, but the average improvement is small because the adaptive sample already exposes the main recurring error mechanisms. This supports the design goal of collecting enough labels for reliable synthesis without turning development-set construction into full-table annotation.

The random 1\% setting is much less stable. Its average F1 falls to 0.77, driven mainly by \textit{hospital}, where only ten labeled rows expose too few of the sparse error-bearing columns. In contrast, \textit{tax} remains near perfect under 1\% sampling because 1\% of that table still contains 2,000 rows and many examples of its regular formatting errors. Across settings, precision remains high. Most quality differences come from recall. Thus, insufficient development coverage primarily prevents \system from discovering supported repairs, rather than causing broad over-correction.

Halving the adaptive set reduces average F1 to 0.87 and especially hurts \textit{hospital}, but several datasets remain robust. This indicates that some domains need only a few examples of regular transformations, while heterogeneous schemas benefit from the adaptive coverage checks. Overall, the adaptive policy provides a practical default. It approaches the quality of a doubled sample while using fewer annotations and avoiding the brittleness of fixed-percentage sampling.

\subsection{RQ6: Model Sensitivity}
\label{sec:model-sensitivity}

\begin{figure}[t]
\centering
\includegraphics[width=0.92\columnwidth]{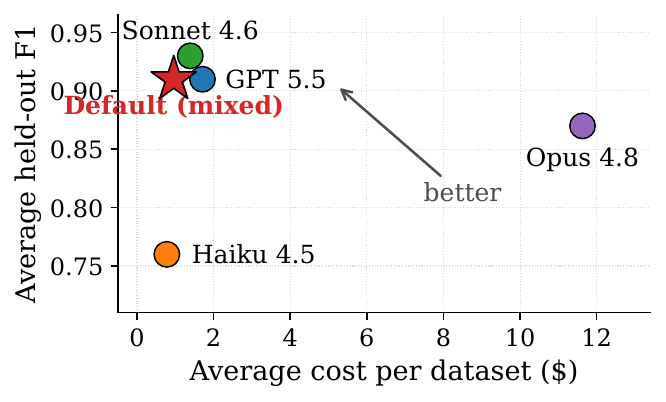}
\caption{RQ6 model sensitivity. Average cost per dataset vs.\ average held-out F1 when a single model drives all agents. Up and to the left is better. The star marks the default mixed-model configuration.}
\label{tab:model-sensitivity}
\end{figure}

We evaluate how different LLMs affect cleaning quality and cost when the same model is used for all agents. We compare GPT 5.5, Opus 4.8, Sonnet 4.6, and Haiku 4.5 on the same six datasets. Figure~\ref{tab:model-sensitivity} plots each model's average held-out F1 against its average monetary cost per dataset.

Figure~\ref{tab:model-sensitivity} shows that model choice changes both average quality and dataset-specific behavior. Sonnet 4.6 gives the best overall result, reaching 0.93 average F1 and matching or improving the default configuration on six datasets. GPT 5.5 is also strong, with 0.91 average F1 and the best \textit{hospital} result, but it is weaker on \textit{flight} than Sonnet 4.6 and Opus 4.8. These results suggest that a capable single-model configuration can simplify deployment, but the default mixed-model configuration remains competitive.

Opus 4.8 does not dominate despite its higher cost. It performs well on \textit{flight}, \textit{beers}, \textit{tax}, \textit{rayyan}, and \textit{movies}, but drops to 0.62 F1 on \textit{hospital}. Its average cost is also much higher, about \$11.63 per dataset compared with \$1.39 for Sonnet 4.6, \$1.71 for GPT 5.5, and \$0.96 for the default mixed-model setting. The expensive model therefore provides neither the best mean quality nor the best cost-quality tradeoff in this workflow.

Haiku 4.5 is the cheapest single-model option but is less reliable. It remains competitive on highly regular transformations such as \textit{beers}, \textit{tax}, and \textit{movies}, yet fails to recover most repairs on \textit{rayyan}, where F1 falls to 0.17. The failure is mostly recall-driven. The generated programs make few false positives but miss many supported fixes. This reinforces a broader pattern from RQ4 and RQ5. The workflow benefits from high-precision validation, but recall depends on the model's ability to infer across refinement rounds.

\subsection{Generated Program Case Study}
\label{sec:program-case-study}
Figure~\ref{fig:program-case-study} shows a shortened excerpt from the actual best program synthesized for the \textit{tax} benchmark. The diagnosis and strategy artifacts identified two high-confidence format errors from the annotated development rows. Integer-valued tax rates were represented as floating strings, such as \texttt{7.0} or \texttt{10.00}, and some ZIP codes were represented with leading zeros, such as \texttt{00627}. The clean target was not a learned label lookup. It was a schema-level transformation that removes the zero-valued fractional suffix for \texttt{rate} and strips leading zeros for digit-only \texttt{zip} values.

\begin{figure}[h]
\centering
\begin{lstlisting}[style=tabclean]
(*@\textcolor{tcDirty}{\ttfamily\detokenize{rate_dirty_re = re.compile(r'^[0-9]+\.0+$')}}@*)
(*@\textcolor{tcDirty}{\ttfamily \# matches dirty whole-number floats}@*)
(*@\textcolor{tcGuard}{\ttfamily\detokenize{rate_clean_re = re.compile(r'^[0-9]+$')}}@*)
(*@\textcolor{tcGuard}{\ttfamily \# recognizes already-clean integers}@*)
(*@\textcolor{tcGuard}{\ttfamily\detokenize{zip_digits_re = re.compile(r'^[0-9]+$')}}@*)
(*@\textcolor{tcGuard}{\ttfamily \# limits zip repair to digit-only cells}@*)

for idx, val in df["rate"].items():
    s = val.strip()
    (*@\textcolor{tcDirty}{\ttfamily \# dirty pattern: "7.0", "10.00"}@*)
    (*@\textcolor{tcGuard}{\ttfamily \# guard: exact whole-number float, not clean int}@*)
    if s and rate_dirty_re.fullmatch(s) \
         and not rate_clean_re.fullmatch(s):
        (*@\textcolor{tcTransform}{\ttfamily df.at[idx, "rate"] = s.split(".", 1)[0]}@*)

for idx, val in df["zip"].items():
    s = val.strip()
    (*@\textcolor{tcDirty}{\ttfamily \# dirty pattern: leading-zero digit string}@*)
    (*@\textcolor{tcGuard}{\ttfamily \# guard: pure digits, starts with 0, nonempty result}@*)
    if s and zip_digits_re.fullmatch(s) \
         and s.startswith("0"):
        stripped = s.lstrip("0")
        if stripped:
            (*@\textcolor{tcTransform}{\ttfamily df.at[idx, "zip"] = stripped}@*)
\end{lstlisting}
\caption{Shortened excerpt of a real synthesized \textit{tax} cleaning program. Red comments mark dirty patterns, blue comments mark guard predicates, and green lines mark the clean transformations.}
\label{fig:program-case-study}
\end{figure}

The guards are what make the artifact reusable rather than merely generative. For \texttt{rate}, the program fires only when the trimmed value exactly matches the dirty pattern \texttt{\textasciicircum[0-9]+{\textbackslash}.0+\$}. Already-clean integers such as \texttt{7} fail the dirty-pattern guard, and meaningful decimals such as \texttt{1.9519792} fail the exact whole-number-float guard. For \texttt{zip}, the program fires only on pure digit strings that start with \texttt{0}, and it refuses to overwrite a cell if stripping zeros would produce the empty string. Thus a cell is updated only when it is both pattern-positive and target-negative. Already-clean cells bypass the rule through ordinary control flow, not through another LLM judgment.

This synthesized program also illustrates held-out reuse. It was selected using 150 development rows, then applied unchanged to the held-out portion of the 200,000-row \textit{tax} table, where it achieved 1.00 precision, 0.976 recall, and 0.988 F1 with 118,150 true-positive repairs and zero false positives. Because the code depends only on column names, regular-expression guards, and deterministic string transformations, the same artifact can be inspected by a data owner, committed under a schema version, cached after validation, and run on future \textit{tax} batches without additional LLM calls. If a later batch introduces a new error mechanism, the cached program may lose recall, but its existing guarded clauses remain auditable and do not expand their repair scope beyond the validated patterns.

\section{Related Work}
\label{sec:related}

\noindent\textbf{Data cleaning.}
Classical data-cleaning systems rely on integrity constraints, statistical inference, or learned repair models to detect and correct dirty values~\cite{datacleaning,trends,detectingerrors}. Constraint-based methods use functional dependencies, conditional functional dependencies, denial constraints, and related rules to identify violations and infer repairs~\cite{tax,holistic,scared,holoclean}. Learning-based systems reduce manual rule engineering by using labels, weak supervision, transfer learning, or active feedback~\cite{activeclean,raha,holodetect,baran,eracer}. These approaches provide strong foundations for evidence-driven repair, but they typically require dataset-specific constraints, features, labels, or retraining. While \system is complementary, it uses profiles and a small development set to synthesize executable repair logic, rather than requiring the user to fully specify constraints or train a dataset-specific model.

\noindent\textbf{LLM-based data preparation and cleaning.}
Recent work applies LLMs to data wrangling, preprocessing, retrieval-assisted repair, standardization, workflow generation, and iterative cleaning~\cite{llmwrangle,llm_processor,retclean,cleanagent,autodcworkflow,iterclean,gidcl,beaver}. Most systems use LLMs as direct operators over cells, rows, chunks, or repeated workflow steps, which makes model inference part of the recurring cleaning path. Other systems adapt or fine-tune models for particular cleaning tasks, improving task specialization but introducing training-data and transfer learning costs~\cite{gidcl,beaver}. \system instead uses LLMs only during synthesis and refinement. After validation, the generated program becomes the reusable cleaning artifact, so applying it to large or future schema-compatible tables does not introduce additional costs.

\noindent\textbf{Executable cleaning logic and agentic validation.}
The closest line of work uses LLMs to generate cleaning logic rather than directly editing every value. LLMClean generates Ontological Functional Dependencies for context-aware repair~\cite{llmclean}, while Cocoon synthesizes executable transformations such as regular expressions and SQL \texttt{CASE WHEN} clauses~\cite{cocoon}. Beyond repair, Castle compiles LLM reasoning into causally consistent SQL \texttt{UPDATE} statements that propagate an intended change across dependent columns while keeping table content hidden from the model~\cite{su-etal-2025-castle}. \system shares the goal of compiling LLM reasoning into executable artifacts, but differs in scope and control. It synthesizes guarded Python repair programs, validates them with cell-level development-set feedback, and refines them through a finite-state multi-agent workflow. This design follows the broader trend of tool-grounded LLM agents for data management and software engineering, where LLM proposals are checked by deterministic execution and feedback loops~\cite{tutotial,quite,sweagent,selfdebugging,reflexion}. In \system, the controller commits only the best validated program, separating stochastic agent reasoning from deterministic execution, scoring, and stopping.

\section{Discussion and Future Work}
\label{sec}

\system is intentionally conservative. It changes cells only when a repair can be expressed as guarded program logic and validated on the development set. This design helps preserve precision, which is critical for data-cleaning workloads where an incorrect repair may be more harmful than leaving a value unchanged. However, this conservatism also limits recall. \system may miss repairs whose evidence is sparse, external to the table, or semantically ambiguous. The \textit{hospital} results illustrate this limitation, where identifier-like and address-like fields require domain knowledge or external evidence beyond a small annotated sample, so a program synthesized only from table-local examples may be unable to distinguish a valid but rare value from an erroneous one.

Another limitation is that \system currently treats reuse as a guarded application of a previously validated program. This is effective when future batches follow the same schema and error distribution, but real data pipelines may evolve as column meanings drift, formatting conventions change, and new error types appear. Although our guarded programs reduce the risk of applying an over-broad repair, stronger reuse-time checks and lightweight program adaptation are needed before deploying cached cleaning programs in long-running pipelines.

Future work can extend \system in three directions. First, retrieval-backed evidence and domain knowledge bases could support repairs that cannot be inferred from the table alone. For example, external dictionaries, address databases, ontology constraints, or historical clean tables could help validate uncertain repairs and expand the coverage of program synthesis. Second, interactive validation and uncertainty estimates could help users approve high-impact repairs while keeping the automatic program conservative. Rather than asking users to inspect every changed cell, the system could surface only representative or high-uncertainty cases, allowing limited human feedback to guide safer repair programs. Third, stronger reuse checks could detect schema drift and distribution shift before a cached program is applied to future batches. More importantly, such checks could enable low-cost incremental adaptation rather than requiring the system to synthesize a new program from scratch.

\section{Conclusion}
\label{sec:conclusion}

We presented \system, a model-training-free tabular data-cleaning system that uses LLMs to synthesize reusable cleaning programs rather than directly repairing individual cells. \system turns a small development set into structured evidence, uses role-specialized agents to diagnose errors and plan guarded repairs, and validates each generated program through deterministic execution and cell-level feedback. The final artifact is an inspectable and cacheable Python program that can be applied to large tables and reused on future schema-compatible batches without additional LLM inference.

Across six standard benchmarks, \system achieves high precision and improves F1 over the
state-of-the-art baselines on five datasets. \system suggests that reusable program synthesis is a promising abstraction for making LLM-assisted data cleaning more scalable, auditable, and practical in real data pipelines.

\balance
\bibliographystyle{IEEEtran}
\bibliography{sample}

\end{document}